\documentclass[fleqn,usenatbib]{mnras}
\usepackage[T1]{fontenc}

\usepackage{graphicx}	
\usepackage{amsmath}	
\usepackage{amssymb}	
\usepackage{subcaption}
\usepackage{float}
\usepackage{enumerate}
\usepackage{multirow}
\usepackage{tabularx}
\usepackage{soul}

\title[The $\alpha$/Fe of LEGA-C quiescent galaxies]{The elemental abundance of quiescent galaxies in the LEGA-C survey: the (non-)evolution of [$\alpha$/Fe] from $z=0.75$ to $z=0$.}

\author[D. Bevacqua et al.]{Davide Bevacqua,$^{1,2}$\thanks{E-mail: davide.bevacqua@inaf.it}, Paolo Saracco$^{1}$, Francesco La Barbera$^{3}$, Giuseppe D'Ago$^{4}$, Roberto De Propris$^{5,6}$, \newauthor Ignacio Ferreras$^{7,8,9}$, Anna Gallazzi$^{10}$, Anna Pasquali$^{11}$, Chiara Spiniello$^{3,12}$
\\
$^{1}$INAF - Osservatorio Astronomico di Brera, via Brera 28, 20121 Milano, Italy\\
$^{2}$DiSAT, Universit\'{a} degli Studi dell'Insubria, via Valleggio 11, I-22100 Como, Italy\\
$^{3}$INAF -  Osservatorio Astronomico di Capodimonte, Via Moiariello  16, 80131, Naples, Italy\\
$^{4}$Instituto de Astrof\'{i}sica, Pontificia Universidad Cat\'{o}lica de Chile, Av. Vicu\~{n}a Mackenna 4860, 7820436 Macul, Santiago, Chile\\
$^{5}$FINCA, University of Turku, Vesilinnantie 5, Turku, 20014, Finland\\
$^{6}$Department of Physics and Astronomy, Botswana International University of Science and Technology, Private Bag 16, Palapye, Botswana\\
$^{7}$Instituto de Astrofisica de Canarias, Calle V\'{i}a Lactea s/n, E38205, La Laguna, Tenerife, Spain\\
$^{8}$ Department of Physics and Astronomy, University College London, London WC1E 6BT, UK\\
$^{9}$Departamento de Astrofisica, Universidad de La Laguna, E38206 La Laguna, Tenerife, Spain\\
$^{10}$INAF - Osservatorio Astrofisico di Arcetri, Largo Enrico Fermi5, I-50125 Firenze, Italy\\
$^{11}$ Astronomisches Rechen-Institut, Zentrum f\"{u}r Astronomie der Universit{\"a}t Heidelberg, M\"{o}nchhofstrasse 12 - 14, 69120 Heidelberg, Germany\\
$^{12}$Sub-Dep. of Astrophysics, Dep. of Physics, University of Oxford, Denys Wilkinson Building, Keble Road, Oxford OX1 3RH, UK\\
}

\date{Accepted 2023 July 27. Received 2023 July 27; in original form 2023 January 16}

\pubyear{2023}

\begin{document}
\label{firstpage}
\pagerange{\pageref{firstpage}--\pageref{lastpage}}
\maketitle

\begin{abstract}
We measure the [$\alpha$/Fe] abundances for 183 quiescent galaxies at $z=0.60-0.75$ with stellar masses ranging $10.4 \leq \log_{10}$(M$_*$/M$_\odot) \leq 11.6$ selected from the LEGA-C survey. We estimate [$\alpha$/Fe] from the ratio of the spectral indices Mgb ($\lambda \sim 5177$ \AA) and Fe4383, compared to predictions of simple stellar population models. We find that $91\%$ of quiescents in our sample have supersolar [$\alpha$/Fe], with an average value of [$\alpha$/Fe] = $+0.24\pm0.01$. We find no significant correlation between [$\alpha$/Fe] and stellar metallicity, mass, velocity dispersion, and average formation time. Galaxies that formed the bulk of their stellar mass on time scales shorter than 1 Gyr follow the same [$\alpha$/Fe] distribution as those which formed on longer time scales. In comparison to local early-type galaxies and to stacked spectra of quiescent galaxies at $z=0.38$ and $z=0.07$, we find that the average [$\alpha$/Fe] has not changed between $z=0.75$ and the present time. Our work shows that the vast majority of massive quiescent galaxies at $z\sim0.7$ are $\alpha$-enhanced, and that no detectable evolution of the average [$\alpha$/Fe] has taken place over the last $\sim$6.5 Gyr.
\end{abstract}




\section{Introduction}

The elemental abundances of galaxies hold fundamental information about their formation history. In particular, metallicity plays a major role, since it is primarily driven by stellar nucleosynthesis, and it is thus tightly related to the stellar content and the star formation history (SFH) of galaxies. This latter is firstly built up by the gas accreted from the intergalactic medium (IGM), which is metal-poor, and subsequently by the evolved stellar populations, which predominantly determine the metallicity of galaxies. 

The interstellar medium (ISM) of galaxies is mostly enriched by supernova (SN) explosions and by stellar winds from stars in the asymptotic giant branch. In particular, ejecta from SNe Type II mostly enrich the ISM with $\alpha$ elements, like O, Ne, Mg, and Si; on the other hand, ejecta from SNe Type Ia mostly produce iron (Fe) peak elements. The explosion timescales for the former are very short, being the consequence of the core collapse of massive stars, while the latter takes longer to explode, being the consequence of the evolution of a white dwarf star, in a binary system, exceeding its limit mass.

At the earliest time of its formation, after the initial stellar burst, a galaxy starts increasing its metallicity due to SNe II, thus enriching the ISM primarily with $\alpha$-elements. Afterward, lower mass stars also evolve and SNe Ia start exploding, increasing the Fe abundance and thus reducing the [$\alpha$/Fe] content of the ISM. As a consequence, the newly formed stars, and thus the average stellar content of the galaxy, will have lower [$\alpha$/Fe]. This implies that galaxies that formed with shorter star formation timescales would have a higher [$\alpha$/Fe] (e.g. \cite{Matteucci_87, Thomas+99}). 

The overall enrichment depends on the amount of metals returned to the ISM and on the capability of a galaxy to retain these metals, and reprocess them into new generations of stars. The global metallicity of stars thus increases when the enriched ISM forms new stars. Further gas accretion from the IGM, instead, dilutes the metallicity of the ISM and enhances star formation, thus reducing the average stellar metallicity.

Studying the relative abundances of $\alpha$ elements and iron elements provides crucial information about the timescales involved in the formation of a galaxy \citep{Tinsley_79, Thomas+99}. In particular, this is a powerful tool to study the formation and evolution of quiescent galaxies, i.e. galaxies that are no longer forming stars. However, a varying stellar Initial Mass Function (IMF) can produce very different [$\alpha$/Fe] even for similar star formation time scales.

In local early-type galaxies (ETGs) with stellar masses M$_* > 10^{10.5} $ M$_\odot$, [$\alpha$/Fe] is observed to increase with stellar mass and velocity dispersion \citep{Jorgensen_99, Trager+00_1, Trager+00_2, Thomas+2005, Gallazzi+2006, Graves+2009a, Graves+2009b, Conroy+2014, LaBarbera+2014, Walcher+2015, Gallazzi+20}, supporting the so-called `downsizing' scenario \citep{Cowie+96,Thomas+10}, where the most massive galaxies formed earlier and faster: their ISM was therefore primarily polluted by core-collapse SNe. More explicitly, due to the short time scales, most stars in massive galaxies formed when the ISM was enriched in $\alpha$ element by the SNe II of the first massive stars, and still not contaminated by iron elements from lower mass stars \citep{Trager+00_1, Conroy+2014}.

While there have been several studies measuring the metallicity of high redshift galaxies (\citealt{Gallazzi+14,Barone+22, ppxf_2022, Carnall+22, EC+19, Kriek+19, Kriek+16, Saracco+23}; see also Bevacqua et al. in preparation), even up to $z\sim3$ \citep{Saracco+20}, only a few studies have concerned the [$\alpha$/Fe] abundances \citep{Choi+14, Onodera+15, Leethochawalit+18, Leethochawalit_2019,Beverage+21, Jafa+20, Saracco+23, Beverage+23}. In this paper, we study the [$\alpha$/Fe] of a sample of quiescent galaxies at redshift $0.6 \leq z \leq 0.75$, and compare it to galaxies in the local Universe, to understand whether their [$\alpha$/Fe] has changed over the last $\sim 6.5$ Gyr of cosmic evolution. Additionally, we attempt to recover the correlations of [$\alpha$/Fe] with galaxies' properties observed at $z \sim 0$.


The paper is organized as follows. In section \ref{sect:data} we describe the data used for the analysis, and the criteria used to build our galaxy sample. Then, in section \ref{sect:stellar_pop} we describe the fitting method used to derive the stellar population properties of our galaxy sample. Since we retrieve estimates of [$\alpha$/Fe] from spectral indices, we also evaluate the impact of age and metallicity on the indices used for the analysis. In section \ref{sect:alpha/fe} we describe how we estimate [$\alpha$/Fe] from the observed spectral indices. Then, in section \ref{sect:alpha_evol} we probe the evolution of the [$\alpha$/Fe] by comparing LEGA-C galaxies with galaxies at lower redshift. Finally, in section \ref{sect:summary} we summarize the results and present our conclusions.

This work is the second part of a study on the overall metallicity of quiescent galaxies at intermediate redshift. We refer to the other paper, Bevacqua et al. (in prep.) as Paper I.

In this paper we adopt a flat $\Lambda$CDM cosmology with H$_0 = 70$ km s$^{-1}$ and $\Omega_{\rm m} = 0.3$.
\section{Data and sample selection}\label{sect:data}

In this paper, we analyze a sample of quiescent galaxies selected from the Large Early Galaxy Astrophysics Census survey (\cite{legac}, hereafter, LEGA-C). LEGA-C is a spectroscopic survey of galaxies at $z \sim 0.6 - 1.0$ in the COSMOS field \citep{cosmos}, using the VIsible Multi-Object Spectrograph (VIMOS; \citealt{vimos}) on the Very Large Telescope. The survey observed 4209 galaxies, selected from the UltraVISTA catalog \citep{ultravista}, reaching an approximate signal-to-noise (S/N) of about 20 \AA$^{-1}$ in the continuum. We use integrated spectra from the third Data Release (\citealt{DR3}; hereafter, DR3), with a nominal spectral resolution of R $\approx$ 2500 \citep{Straatman+18} and observed wavelength range $6300 \mbox{\AA} - 8800 \mbox{\AA}$.

The sample of quiescent galaxies selected from LEGA-C is presented in Paper I. Here, we briefly summarize the selection procedure, and describe the subsample studied here.

We first select quiescent galaxies according to the empirical diagnostic of the UVJ diagram; then, we exclude problematic galaxies using the LEGA-C flags\footnote{We exclude galaxies with FLAG\_MORPH = 1, FLAG\_MORPH = 2 and FLAG\_SPEC = 2.}; finally, we remove galaxies with axial ratio q$_{\rm ax}$ < 0.3 and S\'{e}rsic index n$_{\rm sers}$ < 2.5 to minimize (maximize) the possible presence of late-(early-)type galaxies. The sample is then composed of 637 quiescent galaxies in the redshift range $0.6 \leq z \leq 1$. 

From this sample, we extract a subsample suitable to study [$\alpha$/Fe], according to the following criteria. The best tracer for the abundance of the so-called $\alpha$-elements is magnesium (Mg) which is better traced by the Mgb (5177\AA) index (e.g., \citealt{Thomas+2003}). Given the limited spectral range of LEGA-C spectra, the highest redshift at which Mgb falls in the observed wavelength range is $z\simeq0.75$. Furthermore, some spectra at lower redshift may not have Mgb measured by the DR3, even though the spectral coverage would include the line. This depends on how the DR3 has handled possible systematic effects (like bias in the sky subtraction, bad wavelength calibration, etc.) in the LEGA-C spectra (see section 3.4 of \citealt{DR3} for details). Therefore, we restrict our analysis to those galaxies with measurements of Mgb, thus reducing the initial sample of quiescent galaxies to a subsample of 183 quiescents, within the redshift range $0.6 \leq z \leq 0.75$.

\begin{figure*}
\includegraphics[width=\textwidth]{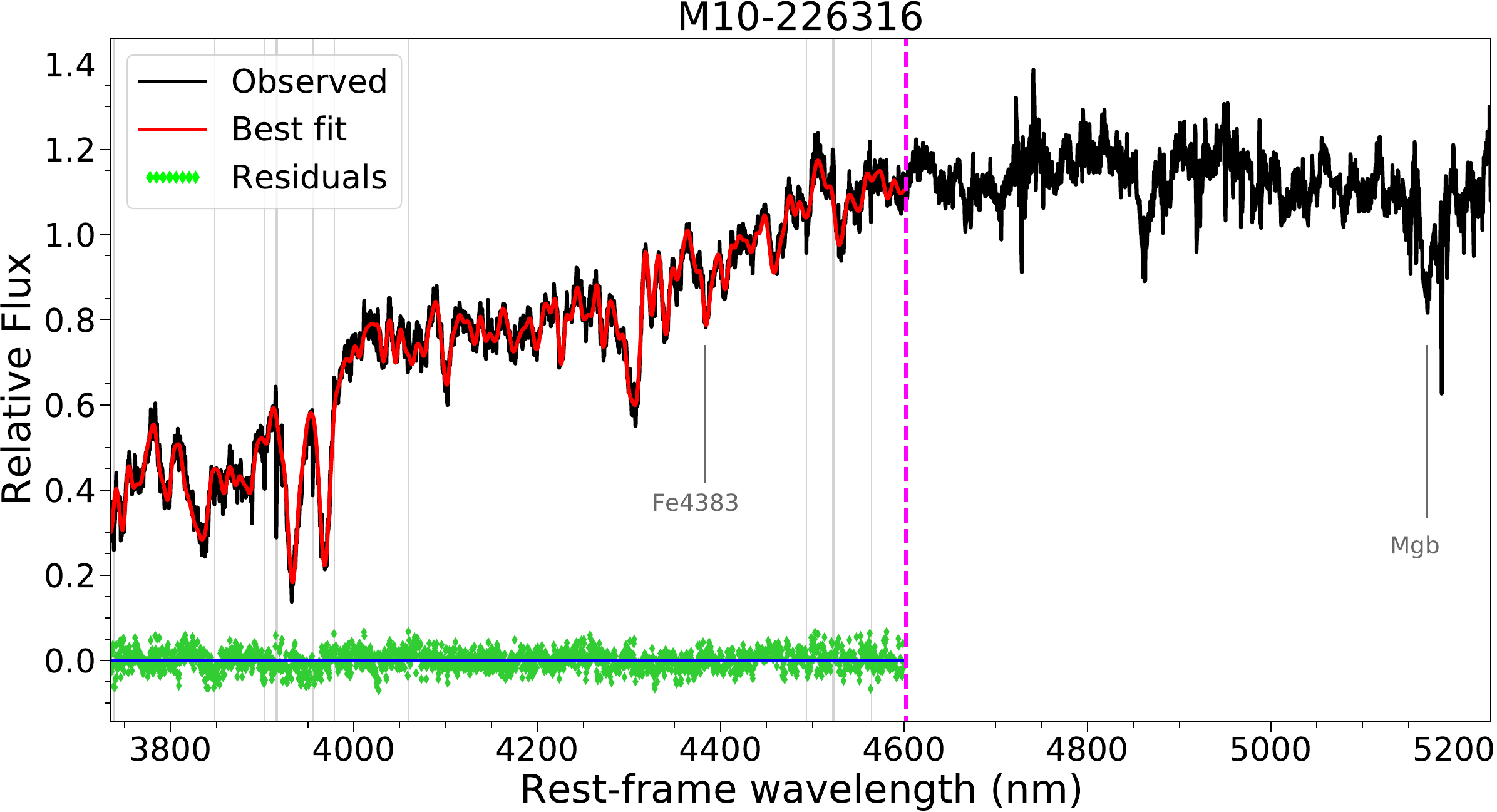}
\caption{Example of a spectrum of a quiescent galaxy from our sample (LEGA-C ID: M10-226316). The black line is the observed spectrum. The red line is the best-fit model. The green diamonds are the residuals, whose median value is indicated by the blue horizontal line. The grey-shaded lines are the masked regions. The dashed magenta line delimits the fitted spectral region (3600 - 4600 \AA ). The spectral indices Fe4383 and Mgb are highlighted in grey at the corresponding wavelengths.}
\label{fig:fitexample}
\end{figure*}

The DR3 of LEGA-C also provides measurements of Mg1 (5102\AA) and Mg2 (5175\AA). However, while Mgb virtually traces only the magnesium abundance, Mg1 and Mg2 are `contaminated' by other metals (primarily Carbon, \citealt{Bernardi+98, Bernardi+06, Thomas+2011, Johansson+2012}).

As a proxy for the iron (Fe) abundance, we use the spectral index Fe4383, which is measured for all galaxies with available Mgb. The other iron lines observed in LEGA-C spectra (Fe4531, Fe5015, Fe5270, Fe5335, and Fe5406) generally have large uncertainties or are not always measured, often falling outside the covered spectral range, thus significantly reducing the statistics. For completeness, in appendix \ref{app:fe} we show that including other iron lines to estimate the [$\alpha$/Fe] does not change our results.

As a consistency test, we have measured ourselves the Mgb and Fe4383 indices for the selected galaxies using LECTOR\footnote{Available from \protect{\url{http://research.iac.es/galeria/vazdekis//vazdekis_software.html}.}}, and compared with the values provided by the DR3. In general, we find a good agreement. Fe4383 and Mgb show scatters of 0.15 and 0.23 around the one-to-one relation, respectively, both largely consistent with the corresponding typical errors, 0.63 and 0.55. For our analysis, we though use the Fe4383 and Mgb indices provided by LEGA-C DR3 being them publicly available and corrected for deviating, high-noise wavelength elements.

Finally, we verified that, in general, the Mgb and Fe4383 indices are not significantly affected by telluric lines. Specifically, we considered those galaxies possibly affected by the most prominent telluric lines (8345 \AA, 8827 \AA, and 8886 \AA), and compared with those which are not affected. We find that the two subsamples follow the same distribution on the Mgb-Fe4383 plane, as confirmed by the 2d KS-test (p-value $\approx 0.4$).

To summarize, in this paper, we study a sample of 183 quiescent galaxies selected from LEGA-C, in the redshift range $0.60 \leq z \leq 0.75$. Stellar velocity dispersion ($\sigma_*$) values are taken from the DR3 of LEGA-C. We estimate the stellar masses by fitting the UltraVISTA photometry \citep{ultravista} with the spectroscopic redshifts provided by LEGA-C, using the C++ implementation of the \texttt{FAST} code\footnote{Available from \url{https://github.com/cschreib/fastpp}} \citep{fast}. To perform the fit, we use models by \cite{BC03}, assuming a delayed exponentially declining SFH, the \cite{Chabrier_IMF} IMF, a \cite{KC13} dust attenuation law, and solar metallicity. The Mgb and Fe4383 indices of the stacked spectra presented in section \ref{sect:alpha_stack_comp} are measured using LECTOR.

\section{Stellar Population Properties}\label{sect:stellar_pop}

In this work, we use the stellar mass-weighted age and metallicity values estimated in Paper I. In this section, we first review the fitting method (section \ref{sect:fitting_method}) used to derive these properties. Then, in section \ref{sect:agemet}, we show the age and metallicity values estimated for the subsample studied in this work, and discuss the impact of redshift evolution, from $z=0.75$ to $z=0.60$, on the considered indices.

\subsection{Fitting method}\label{sect:fitting_method}

To estimate the mass-weighted ages and metallicities, we fit the LEGA-C spectra using the penalized pixel fitting (\texttt{pPXF}) method and code described in \citet[updated to version v8.2.2]{ppxf_2004, ppxf_2017, ppxf_2022}. The code performs a full-spectral fit by linearly combining template spectra of given ages and metallicities and then assigning to each template a weight, thus providing a final composite best-fit model, which is the one minimizing the $\chi^2$. The details of the fitting procedure we adopt, and the numerous tests performed to check our results are presented in Paper I. Here, we review our fitting procedure.

As templates, we use the E-MILES Simple Stellar Population (SSP) models \citep{EMILES}, which are entirely based on observed stars. More specifically we use models with BaSTI isochrones \citep{basti} and a Chabrier initial mass function \citep{Chabrier_IMF}. We restrict to the safe ranges described in \cite{EMILES}, namely we use only models with metallicities [M/H] $= -1.79, -1.49, -1.26, -0.96, -0.66, -0.35,$ $-0.25, +0.06, +0.15, +0.26$ and ages $\geq 0.11$ Gyr. As the upper limit to the age of the SSPs, we consider the lookback times corresponding to a formation redshift of $z=10$ (i.e., assuming that the first stars started forming about half Gyr after the Big Bang) at $z=0.65$ and $z=0.75$, namely 6.984 Gyr and 6.426 Gyr, for galaxies with redshift $0.60 \leq z < 0.70$ and $0.70 \leq z \leq 0.75$, respectively.

The fits are performed as follows. First, since the templates have a higher resolution than the LEGA-C spectra, we convolve them with a Gaussian kernel to match the full width half maximum (FWHM) of the observed spectrum ($\approx 3$\AA); the kinematic broadening is taken into account during the fit, using the velocity dispersion measured by pPXF. The observed spectrum is de-reshifted to the rest frame. We thus perform two fits: from the first fit, we get the residuals between the galaxy and the best-fitting model, and make a robust estimate of the standard deviation of these residuals, $\sigma_{\rm std}$, which we use to mask all the spectral pixels deviating more than $3\sigma_{\rm std}$. Thus, we perform a second fit, which gives us the final best-fitting model.

Each fit is performed using both multiplicative polynomials of degree 4 and a Calzetti reddening curve \citep{Calzetti_2000}, over the spectral range $3600 - 4600$ \AA \, because it is the range common to most of LEGA-C spectra\footnote{We extensively discuss these choices in the appendix of Paper I.}. In the fit, we also include gas emission lines, modeled as gaussians. In particular, we fit the Balmer series, for which we fix the flux ratio (tie\_balmer = True), and the [OII] doublet. In Figure \ref{fig:fitexample} we show an example of a fit performed on a LEGA-C galaxy.

\subsection{Estimated ages and metallicities}\label{sect:agemet}

The mean mass-weighted age and metallicity are obtained as weighted averages, calculated as:

\begin{equation}\label{eq:age}
{\rm log}_{10}{\rm Age} = \frac{\sum_i  w_i {\rm{log_{10} Age}}_i}{\sum_i w_i}
\end{equation}

\begin{equation}\label{eq:met}
{\rm [M/H]} = \frac{\sum_i w_i {\rm [M/H]}_i}{\sum_i  w_i}
\end{equation}

\noindent where $w_i$ is the weight of the $i$-th template assigned to the best-fit model, and the sums are performed over all the templates used in the fit.

A detailed estimate of uncertainties is described in Paper I\footnote{Briefly, we take a subsample of galaxies with different S/N, and perform a number of realizations of each spectrum, by shuffling the noise. For each realization, we perform the fit as described in section \ref{sect:fitting_method} and estimate the age and metallicity. Then, as the typical errors on age and metallicity of each spectrum we take the standard deviations of all the realizations and finally compare these errors with the S/N.}. Since we find no clear correlation between errors and S/N, we assume a typical error of 0.07 dex for ages and 0.06 dex for metallicities, corresponding to the median errors of the whole subsample.
\begin{figure}
\includegraphics[width=\columnwidth]{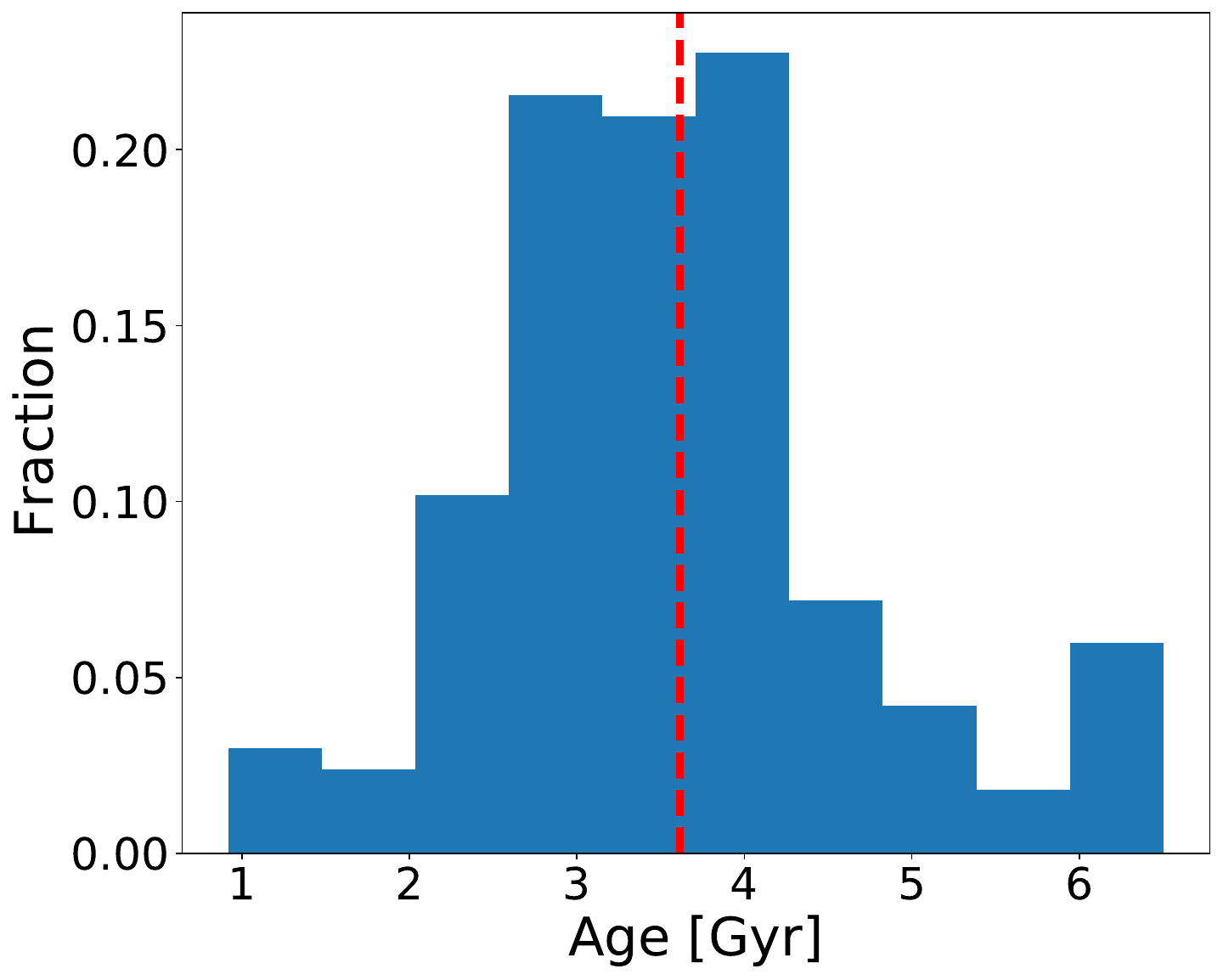}
\includegraphics[width=\columnwidth]{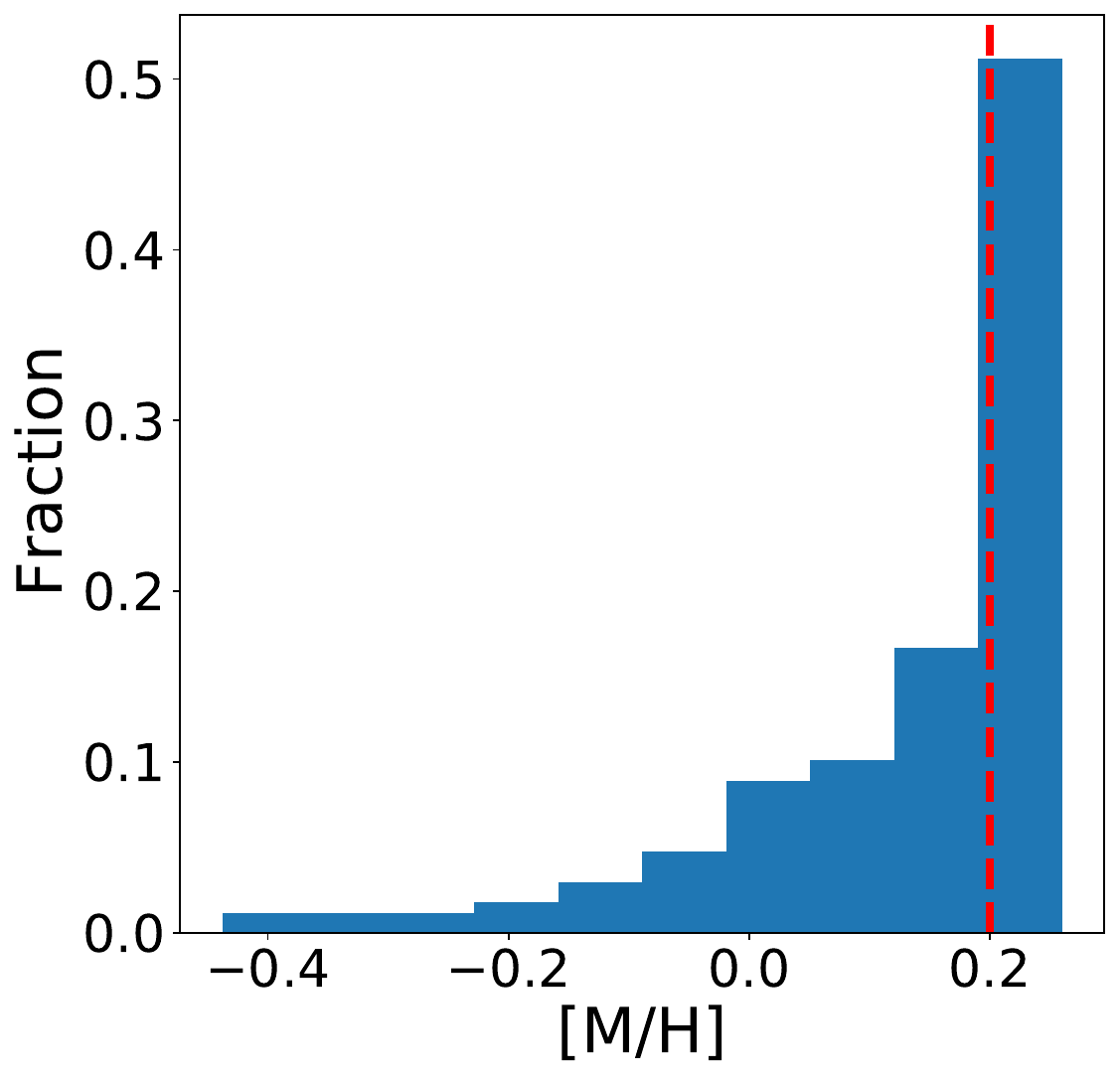}
\caption{Age (upper panel) and metallicity (lower panel) distributions as estimated from the fits. The red dashed line in the Age distribution is the mean value, $\sim$3.6 Gyr, while in the [M/H] distribution is the median value, $\sim$0.2 dex.}
\label{fig:agemet}
\end{figure}

In Figure \ref{fig:agemet} we show the histograms of mass-weighted stellar ages and metallicities estimated from the fits. The distribution of ages is approximately gaussian, with a mean value of $\sim 3.6$ Gyr and a standard deviation of about 1 Gyr. 

We define the cosmic formation time corresponding to the mass-weighted age of a galaxy as:
\begin{equation}\label{eq:tform}
	t_{\rm form} \equiv \mbox{Age}_{\rm U} (z) - \mbox{Age}_{\rm U} (z = 10) - \mbox{Age \; ,}  
\end{equation}
where Age$_{\rm U} (z)$ is the age of the Universe at the redshift at which a galaxy is observed, and Age is the age of the galaxy with respect to $z=10$ estimated from the fits (equation \eqref{eq:age}).

Considering the mean age (3.6 Gyr)  and the standard deviation (1 Gyr) of the sample, it follows that, on average, the stellar population in these galaxies has started to form between 2.7 Gyr < $t_{\rm form}$ < 4.7 Gyr.

The metallicity distribution is skewed towards supersolar metallicities, with a median value of $\approx 0.20$ dex, while about $15\%$ of galaxies have [M/H] < 0.

\begin{figure}
\includegraphics[width=\columnwidth]{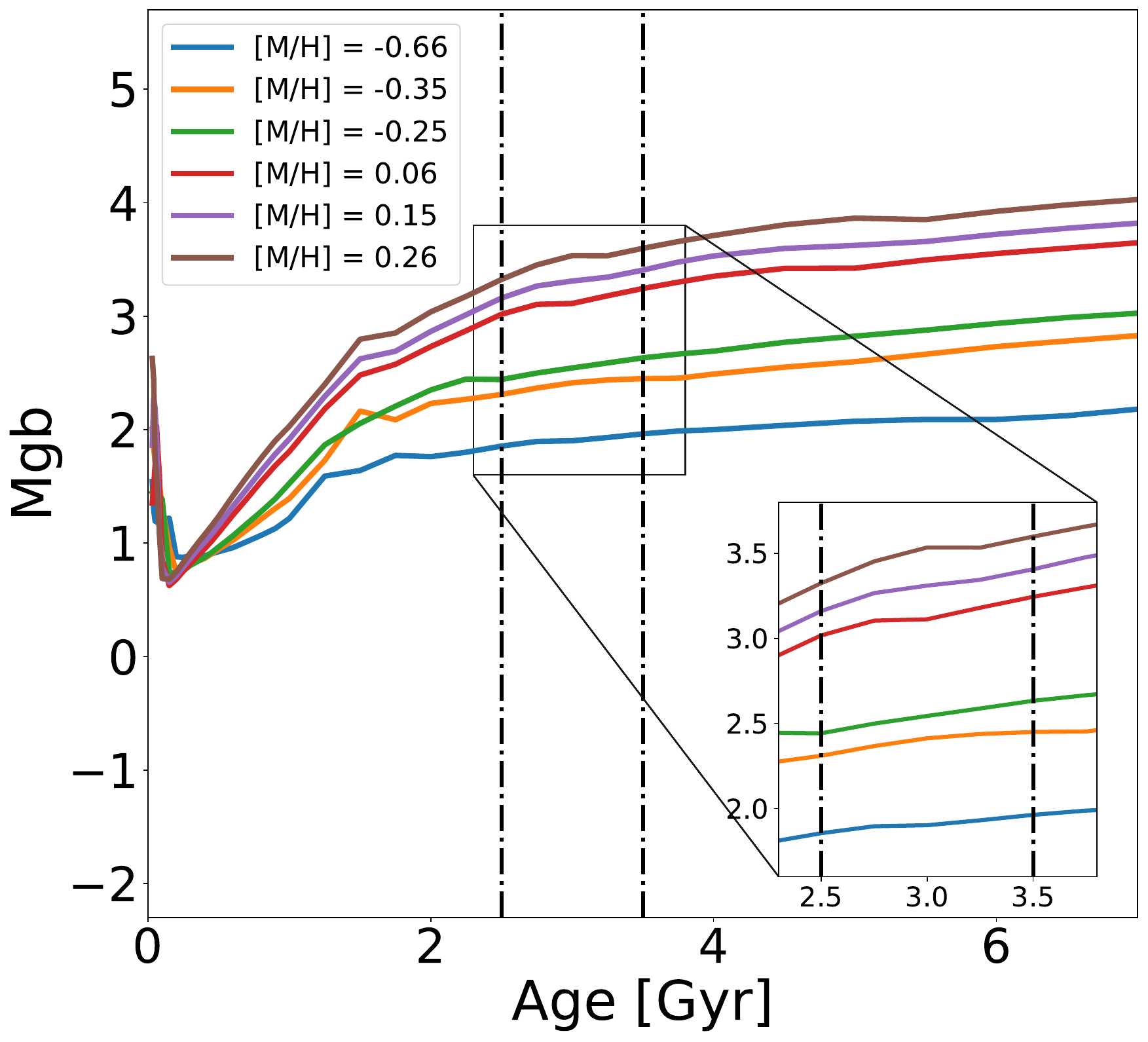}
\caption{Evolution of Mgb as a function of the age, at different metallicities. The values of Mgb are measured by the E-MILES models and provided with the templates library. In this plot, we show that if a galaxy with an age of 3.5 Gyr (about the average of our sample), observed at $z=0.60$, were observed at $z=0.75$ (i.e. 1 Gyr earlier, at 2.5 Gyr), the variation of Mgb would be negligible ($\sim 8\%$), at all estimated metallicities, and well within the uncertainties of the observed values. Similar results are obtained for Mg2 and Fe4383.}
\label{fig:mgb_evol}
\end{figure}

Since galaxies in the sample analyzed here are distributed in a redshift range $0.60 \leq z \leq 0.75$, corresponding to almost 1 Gyr of cosmic time, we should in principle take into account the variation of the indices purely due to the cosmic evolution. As an example, in Figure \ref{fig:mgb_evol} we show how Mgb varies as a function of age for different metallicity values, according to E-MILES models.

Over this interval of cosmic time (1 Gyr), for both Mgb and Fe4383 indices we measure a relative variation of $8\%$, which is much smaller than the typical uncertainty on the measured indices. To derive these estimates, we considered two SSPs with ages 3.5 Gyr (i.e. approximately the average age of our sample) and 2.5 Gyr at fixed metallicity [M/H]$=+0.15$. The variation would be even smaller for older ages since model predictions get flatter (see Figure \ref{fig:mgb_evol}). Analogously, considering different metallicities does not change significantly these variations, as also evident from Figure \ref{fig:mgb_evol} for Mgb (the same result is obtained for Fe4383).

We then conclude that, for our galaxies, the effects of redshift evolution on the two indices within the redshift bin considered can be ignored.

\section{The [$\alpha$/Fe] of quiescent galaxies at z = 0.60 - 0.75}\label{sect:alpha/fe}

In this section, we study the [$\alpha$/Fe] abundances of LEGA-C galaxies by comparing the observed spectral indices with those estimated from MILES models. First, in section \ref{sect:mgfe}, we show how galaxies distribute in the Fe4383 vs Mgb diagram. Then, in section \ref{sect:alpha_estimate}, we estimate the [$\alpha$/Fe] values and study their dependence on galaxies' properties. Finally, in section \ref{sect:alpha_SFH}, we compare the distributions of [$\alpha$/Fe] values of galaxies with different SFHs.

\subsection{The Mgb vs Fe4383 diagram}\label{sect:mgfe}

\begin{figure*}
\includegraphics[width=\textwidth]{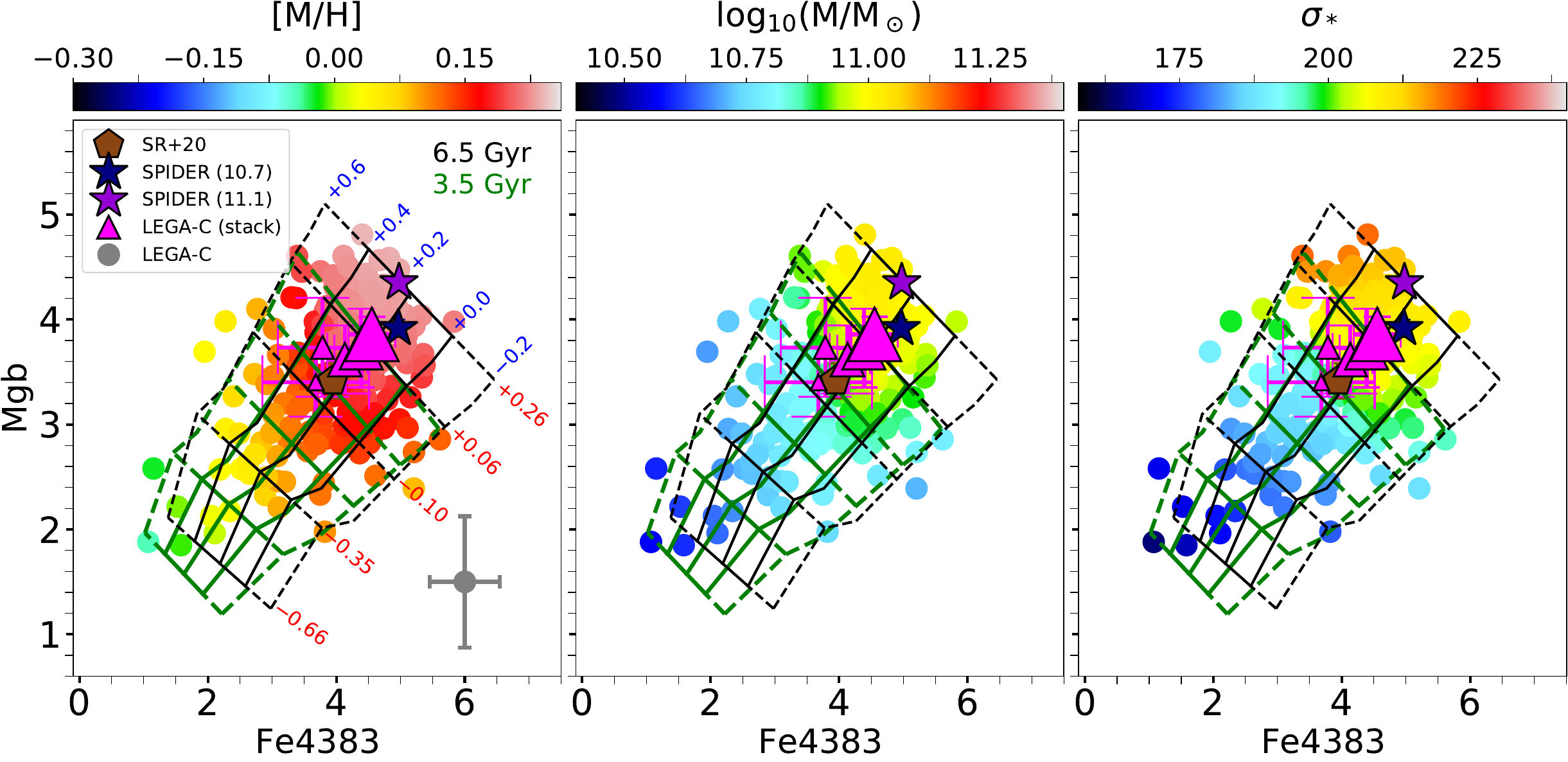}
\caption{Plots of Fe4383 vs Mgb. LEGA-C galaxies are plotted with filled circles. The average errors are plotted in grey in the lower right corners of the two panels of the first column. Colors represent the stellar metallicity (first column), stellar mass (second column), and stellar velocity dispersion (third column); they have been smoothed with \texttt{LOESS} (\citealt{LOESS}, available from \url{https://pypi.org/project/loess/}) using a parameter \texttt{frac = 0.75}. The magenta triangles represent the indices we measure for the stacked spectra of the LEGA-C galaxies, with marker size increasing with the average mass (from 10.5 to 11.3 at step 0.2 in logarithm); errors are the median absolute deviations of the single measurements of LEGA-C galaxies in each stacked spectrum. The navy and violet stars are our measurements of the indices for the stacked galaxies of the SPIDER sample (at $z\sim 0.07$), averaged over masses of 10.7 and 11.1, respectively; errors are the median absolute deviations within the averaged values. Similarly, the brown pentagon is the average value we measure for galaxies from SR+20 (at $z \sim 0.38$), having an average mass of 11.3; errors are the median absolute deviations of the single measurements. The spectral indices measurements of LEGA-C galaxies are taken from the DR3. Instead, we measure Mgb and Fe4383 of all the stacked spectra. The black grid represents the values of Mgb and Fe4383 expected from MILES models of 6.5 Gyr at different values of [$\alpha$/Fe] (blue text in the left panel) and metallicities (red text in the left panel). The green grid is the same but for models with an age of 3.5 Gyr. In both grids, solid (dashed) lines represent the values interpolated (extrapolated) at values of [$\alpha$/Fe] inside (outside) the range considered by the MILES models, i.e. [$\alpha$/Fe] = $0.0$ and $+0.4$.}
\label{fig:mgfe}
\end{figure*}

In Figure \ref{fig:mgfe} we show the Mgb against Fe4383 of our galaxy sample as provided by the LEGA-C DR3. 

To compare the observed indices with models, we measure Mgb and Fe4383 of the MILES models with solar scale [$\alpha$/Fe] and with [$\alpha$/Fe]$=+0.4$, obtained with BASTI isochrones for a Chabrier IMF. In particular, we consider models of fixed age = 3.5 Gyr, corresponding to the average age estimated from the fits of the galaxy spectra, and 6.5 Gyr, corresponding to the oldest age estimated, and at metallicities [M/H] $= -0.66, -0.35, -0.25, +0.06, +0.15, +0.26$. Then, we match the spectral resolution of MILES to that of LEGA-C spectra. To take into account the broadening of the indices due to the stellar velocity dispersion, we convolve models by gaussians with widths $\sigma_* =$ 150, 200, and 250 km s$^{-1}$, corresponding to the 25th, 50th, and 75th percentiles of the distribution in velocity dispersion of our sample. Hence, we build three grids for each age of the models. In Figure \ref{fig:mgfe} we only show the grid convolved by $\sigma_* =$ 200 km s$^{-1}$. However, in the next section, we estimate the [$\alpha$/Fe] of galaxies with  $\sigma_* < 175$ km s$^{-1}$, $175 \leq \sigma_* < 225$ km s$^{-1}$, and $\sigma_* > 225$ km s$^{-1}$ using the grids convolved with the lower, median, and higher velocity dispersions, respectively.

As we discuss in the next section, we provide estimates of the [$\alpha$/Fe] for LEGA-C galaxies by comparing the observed indices with the grids; for this purpose, we here sample the grids using finer steps. To have a finer sampling of [$\alpha$/Fe], for each spectral index we consider the value measured at [$\alpha$/Fe] = 0.0 dex and [$\alpha$/Fe] $= +0.4$ dex, at fixed metallicity. We evenly space this interval at a fixed step, depending on the index considered, and corresponding to a sampling of 0.01 dex in [$\alpha$/Fe]\footnote{For example, at [M/H] = $+0.06$ dex, we measure Mgb $= 3.3666$ and Mgb $= 4.1430$ for [$\alpha$/Fe] = 0.0 and $+0.4$ dex, respectively, corresponding to an interval of 0.776. To have a sampling of 0.01 dex in [$\alpha$/Fe], the corresponding step in Mgb is given by $0.776/[(0.4 - 0.0)/0.01] = 0.019$.}. With the same sampling, we also extrapolate values up to [$\alpha$/Fe] = $+0.6$ dex and down to [$\alpha$/Fe] = $-0.2$ dex, for each spectral index. Similarly, for each re-sampled [$\alpha$/Fe] value, we re-sample the metallicities by evenly spacing models of adjacent metallicities to have 10 intermediate values. This implies that metallicity is not sampled at regular steps, but the sampling depends on the values of adjacent models\footnote{For instance, the step in metallicity between models of [M/H] = $-0.66$ dex and $-0.35$ dex is $(-0.35 + 0.66)/10 = 0.03$ dex while between $+0.15$ dex and $+0.26$ dex is $(0.26 - 0.15)/10 = 0.01$ dex.}. As a final result, for each index, we have two grids of values corresponding to $80 \times 56 $ ([$\alpha$/Fe] $\times$ [M/H]) values, for models of fixed age 6.5 Gyr and 3.5 Gyr. The two grids are shown in Figure \ref{fig:mgfe}, at selected values of [$\alpha$/Fe] and [M/H]. 

Note how the effect of age on the grids is an almost rigid shift, parallel to metallicity, towards higher values of Fe4383 and Mgb at older ages, which does not affect the measurements of [$\alpha$/Fe], at least down to [M/H]$\approx -0.35$ dex. However, even at the lowest metallicities, the single data points are consistent within the uncertainties with the same [$\alpha$/Fe] independently of the grid considered, 3.5 Gyr or 6.5 Gyr.

In Figure \ref{fig:mgfe}, the LEGA-C galaxies are filled circles colored in metallicity, stellar mass, and stellar velocity dispersion. Here, we only show the smoothed values to spot possible correlations with galaxies' properties; we refer to section \ref{sect:alpha_estimate} for a more quantitative study of these trends.

\subsection{Estimates of [$\alpha$/Fe]}\label{sect:alpha_estimate}

\begin{table}
	\centering
	\caption{Table of estimated stellar population parameters}
	\label{tab:alpha}
	\begin{tabularx}{\columnwidth}{>{\centering\arraybackslash}p{2.cm}>{\centering\arraybackslash}p{1.5cm}>{\centering\arraybackslash}p{1.5cm}>{\centering\arraybackslash}p{1.5cm}}
		\hline
		ID LEGA-C & log$_{10}$Age & [M/H] & [$\alpha$/Fe]\\
		 & (yr) & (dex) & (dex)\\
		(1) & (2) & (3) & (4) \\
		\hline
		& & & \\
		M16\_38110 & 9.81 & $-0.10$ & $+0.60^{+ 0.00}_{-0.31 }$ \\
		M14\_41209 & 9.68 & $-0.12$ & $+0.01^{+ 0.39 }_{- 0.21 }$ \\
		M16\_103417 & 9.66 & $+0.13$ & $+0.32^{+ 0.28 }_{- 0.33 }$ \\
		M14\_104576 & 9.54 & $-0.43$ & $+0.55^{+ 0.05 }_{- 0.36 }$ \\
		M7\_108472 & 9.12 & $-0.23$ & $+0.60^{+ 0.00 }_{- 0.23 }$ \\
		... & ... & ... & ...\\
		& & & \\
		\hline
	\end{tabularx}
	\begin{minipage}{\columnwidth}
	List of the first 5 galaxies from our sample of 183 galaxies with measured indices of Mgb and relative stellar population properties. The remaining values are provided as supplementary material to this paper. Columns: (1) ID LEGA-C of the file associated with the spectrum. (2) Mass-weighted age estimated from fits, using equation \eqref{eq:age}; we assume a fixed error of 0.07 dex for all galaxies. (3) Mass-weighted metallicity estimated from fits, using equation \eqref{eq:met}; we assume a fixed error of 0.06 dex for all galaxies. (4) [$\alpha$/Fe] estimated by comparing Mgb and Fe4383 indices with predictions from models; each tabulated value is the median value estimated from 10000 simulations, while the lower and upper uncertainties are the 16th-84th percentiles; see the main text for details.
	\end{minipage}

\end{table}

\begin{figure}
\includegraphics[width=\columnwidth]{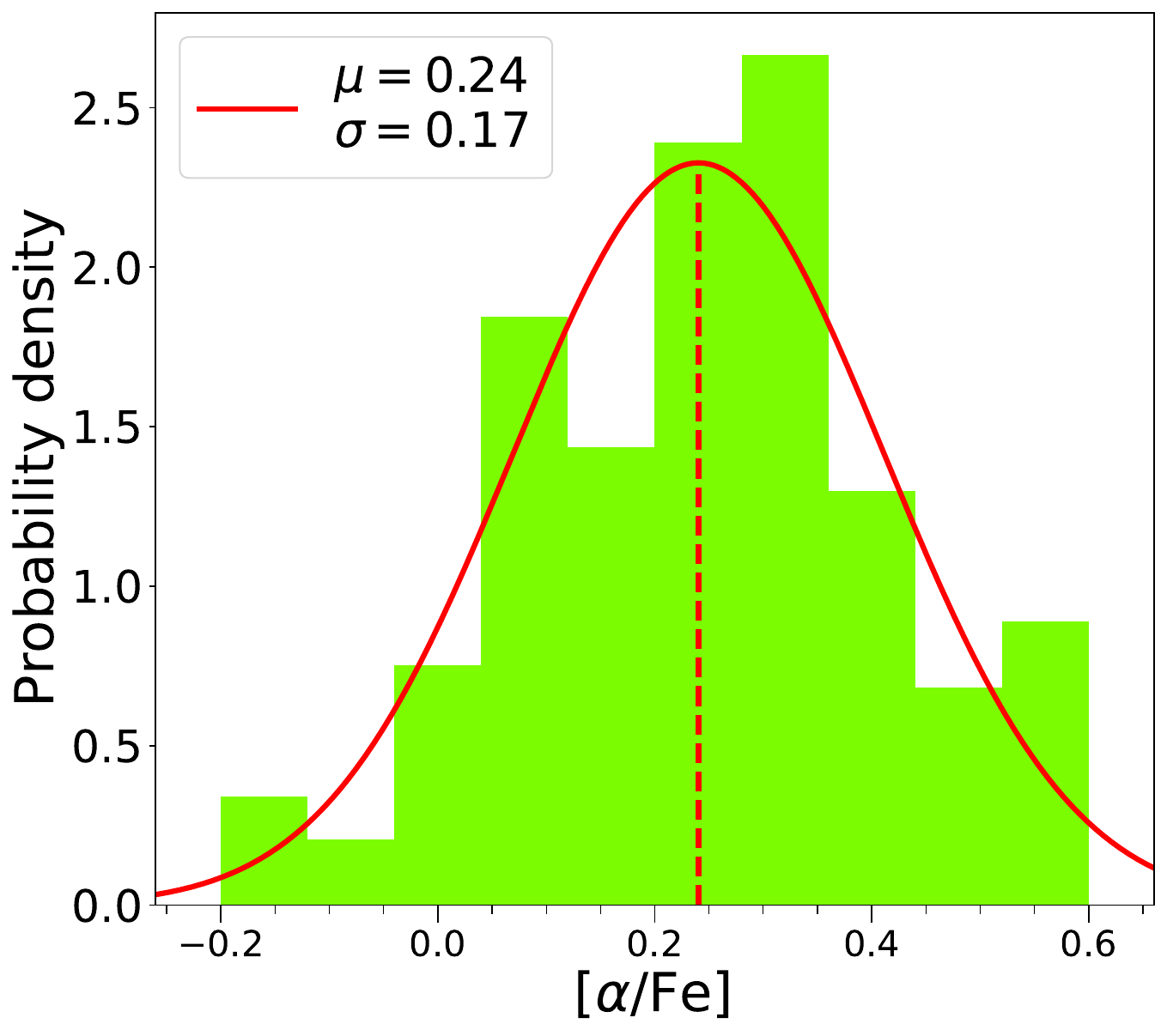}
\caption{Histogram of [$\alpha$/Fe] values as obtained by comparing the Mgb and Fe4383 observed values and the ones estimated from MILES models, using equation \eqref{eq:chi2}. The solid red line represents the gaussian approximation of the histogram, with the dashed lines representing the mean, whose value is shown in the legend along with the standard deviation.}
\label{fig:hist_alpha}
\end{figure}

For each galaxy, we derived [$\alpha$/Fe] by comparing the observed spectral indices with those predicted by the grid of models.

Given the little dependence of the grids on age, we choose to estimate the [$\alpha$/Fe] for all LEGA-C galaxies using the 6.5 Gyr grid, as it provides better coverage of the observed data points, even though most galaxies have ages younger than 6.5 Gyr. \footnote{For those galaxies (14) with low Mgb ($\lesssim 2.5$) and Fe4383 ($\lesssim 4$), whose average age is 3.3 Gyr, we verified that using the 3.5 Gyr grid provides systematically lower [$\alpha$/Fe] values by, on average, $\sim 0.05$ dex, but without affecting significantly the global statistics, and thus our conclusions.} Notice that, unlike [$\alpha$/Fe], the metallicity of the adopted grid depends on age. For this reason, we do not estimate the metallicity from the grid, and rely on the [M/H] estimates from spectral fitting (section \ref{sect:agemet}).

The values of [$\alpha$/Fe] are estimated as follows. For each galaxy, we consider the observed indices I$_{\rm obs}$ (i.e., Mgb and Fe4383) and compute the $\chi^2$, defined as:

\begin{equation}\label{eq:chi2}
\chi^2 = \sum_j \left(\frac{\rm{I}_{\rm{obs},\textit{j}} - \rm{I}_{\rm {mod},\textit{j}}}{\sigma_{\rm{I},\textit{j}}}\right)^2 \; \; \; ,
\end{equation}
\noindent where j runs over the indices Mgb and Fe4383, I$_{\rm{mod}, \textit{j}}$ is the grid (which is a matrix of dimension  [$\alpha$/Fe] $\times$ [M/H] = $80 \times 56 $) of the index considered, and $\sigma_{\rm {I},\textit{j}}$ the corresponding measured error. Then, we take the value of [$\alpha$/Fe] corresponding to the minimum $\chi^2$.


We repeat this procedure $10000$ times, assuming gaussian errors.  We verify that the distributions of these realizations are either gaussian or skewed. In particular, the non-gaussian cases exhibit distributions skewed \textit{towards the extreme values of [$\alpha$/Fe]}, suggesting that galaxies would distribute in a larger range of [$\alpha$/Fe], if available; on the other hand, galaxies with average `central' values of [$\alpha$/Fe] (i.e. about from $+0.1$ dex to $+0.3$ dex) are always gaussian. For this reason, we assign the median value of these realizations as the [$\alpha$/Fe] value for the galaxy, and the 16th and 84th percentiles as the lower and upper errors. In table \ref{tab:alpha} we list 5 galaxies of our sample with the estimated stellar population properties. We provide a machine-readable table with all the estimated values as supplementary material to this paper.

In Figure \ref{fig:hist_alpha} we show the histogram of estimated [$\alpha$/Fe] values. The distribution is fairly gaussian, with a mean of [$\alpha$/Fe] = $+0.24$ and a standard deviation of 0.17, corresponding to a standard error of the mean of $\approx 0.01$ dex. We verified that these results remain virtually unchanged when considering only galaxies with higher S/N. From the plots of Figure \ref{fig:mgfe}, it is already evident that almost all LEGA-C galaxies overlap with models having supersolar [$\alpha$/Fe]. More quantitatively, we find that $91\%$ of galaxies have [$\alpha$/Fe]$> 0$, while only a small fraction ($9\%$) is $\alpha$-depleted or solar-scaled.

\begin{figure}
\includegraphics[width=\columnwidth]{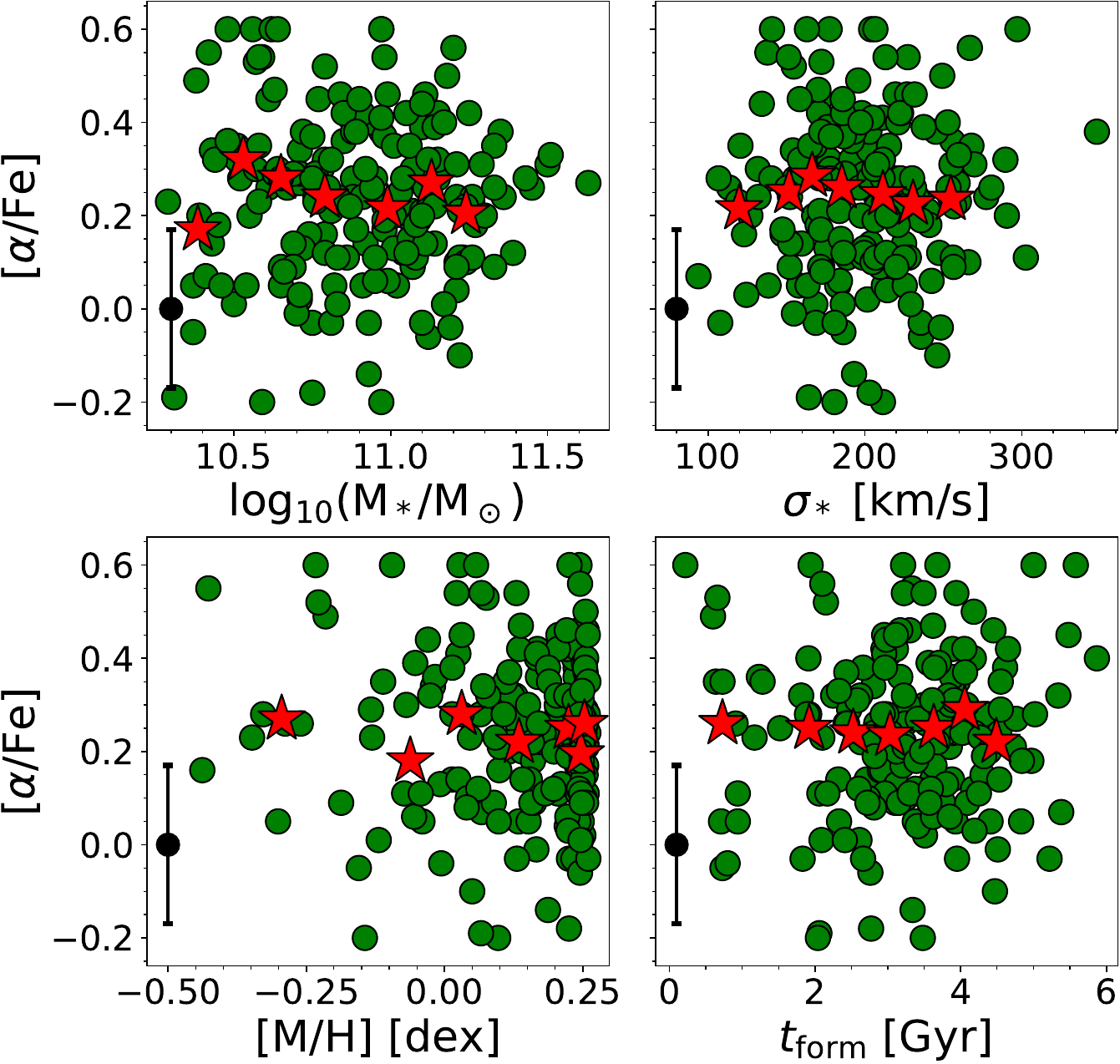}
\caption{Comparison of [$\alpha$/Fe] with stellar mass (top left), stellar velocity dispersion (top right), [M/H] (bottom left), and formation time (bottom right). The red stars are the median [$\alpha$/Fe] at 5th, 16th, 25th, 50th, 75th, 84th, and 95th percentiles of the corresponding x-coordinate. In the bottom left corner of each panel, the typical error on [$\alpha$/Fe] is shown.}
\label{fig:alpha_mms}
\end{figure}

In Figure \ref{fig:alpha_mms} we show the estimated values of [$\alpha$/Fe] as a function of the stellar metallicity, the stellar mass, the stellar velocity dispersion, and formation time (eq. \eqref{eq:tform}). The plots are dominated by the scatter in [$\alpha$/Fe], mainly due to the uncertainties on the indices, so no significant trend is spotted. We verified that no clear trends are found when considering only galaxies with higher S/N. This is in agreement with the lack of correlation between [Mg/Fe] and stellar mass discussed in \cite{Beverage+21} on 82 quiescent galaxies from LEGA-C. In \cite{LaBarbera+2014}, the estimated [$\alpha$/Fe] of SPIDER galaxies varies from $\sim 0.14$ dex to $\sim 0.3$ dex, in a similar range of $\sigma_*$; this variation is of the same order of the average error on [$\alpha$/Fe], i.e. $\sim 0.16$ dex. Also, note that the average error is comparable with the standard deviation of the general [$\alpha$/Fe] distribution  (0.17 dex). Therefore, from these data, we can not confirm the trends found for quiescent galaxies in the local Universe, because of the large uncertainties on measured indices.

\subsection{The distributions of [$\alpha$/Fe] for different SFHs}\label{sect:alpha_SFH}

\begin{figure}
\includegraphics[width=\columnwidth]{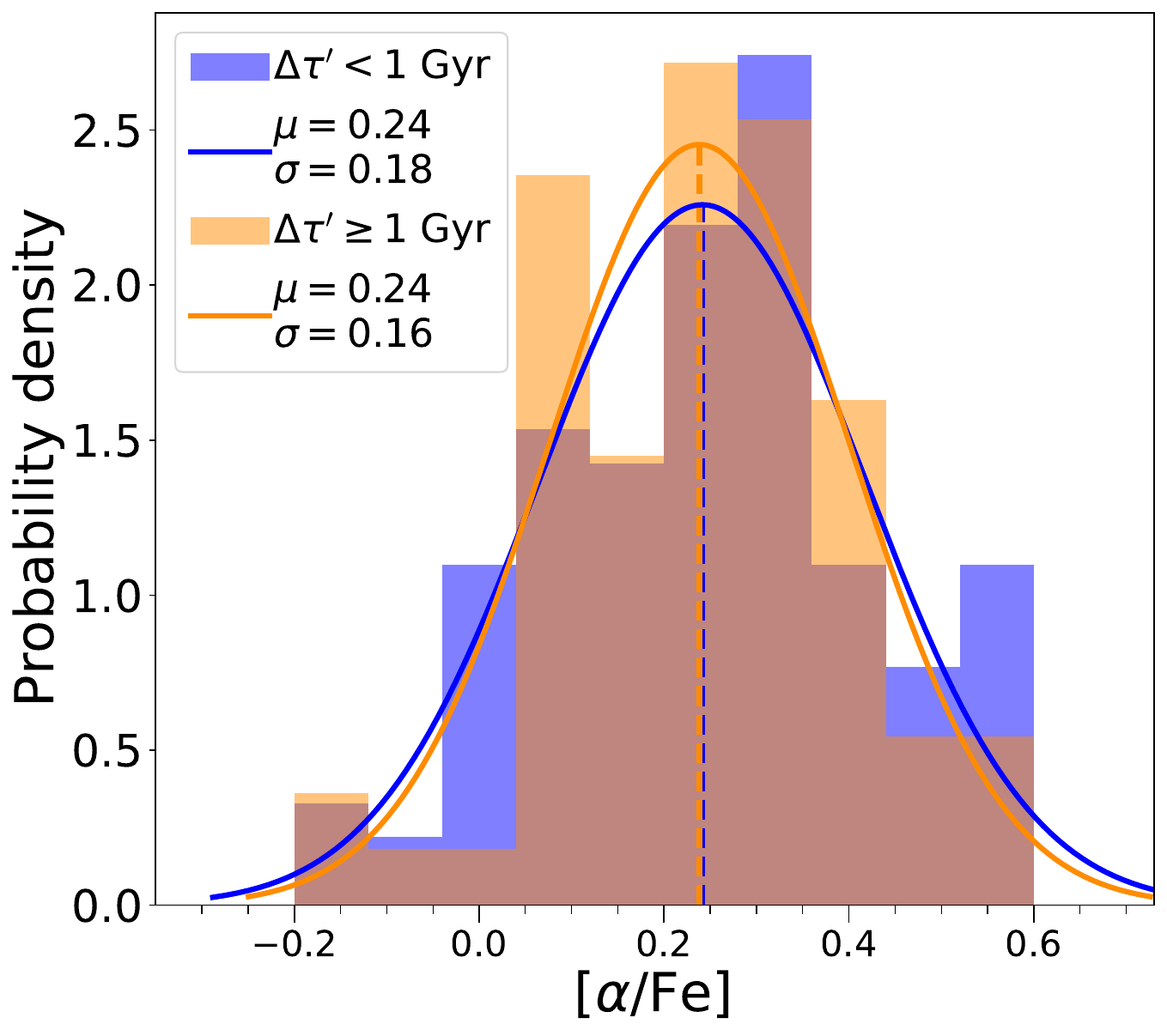}
\caption{Same as Figure \ref{fig:hist_alpha}, but the sample is split in two: galaxies that formed $25\%$ to $75\%$ of their mass in a time $\Delta\tau'$ shorter (blue) or longer (orange) than 1 Gyr. The solid lines represent the gaussian approximation of the two histograms, with the dashed vertical lines representing the relative means, whose values are shown in the legend together with their standard deviations.}
\label{fig:hist_alpha_deltat}
\end{figure}

In Paper I, we study the SFHs of our sample. Briefly, to build the SFH of a galaxy, we consider the mass-weights assigned by \texttt{pPXF} to the input SSPs used to derive the best-fitting spectrum. Then, the SFH is calculated as the cumulative distribution function of the weights as a function of time. For each galaxy, we then define the time, $\tau$, at which a galaxy has reached a certain fraction of the total mass. Here, we consider $\tau_{25}$ and $\tau_{75}$, i.e. the times at which galaxies reach the 25$\%$ and 75$\%$ of their mass, respectively, and compute their difference $\Delta\tau' = \tau_{75} - \tau_{25}$, which gives us a proxy of the formation time scale for the bulk of the stellar mass of galaxies.

We now split the sample in two: galaxies that formed in a $\Delta\tau'$ shorter than 1 Gyr, and galaxies that formed in a $\Delta \tau'$ larger than 1 Gyr. We choose this value, 1 Gyr, because it is the highest uncertainty we have on age, purely due to the sampling in ages of the MILES models (0.5 Gyr, for ages older than 4 Gyr)\footnote{We verified that considering values larger or smaller than 1 Gyr does not change the results.}. This distinction aims at qualitatively dividing galaxies into those for which the largest fraction of stars formed in a very short episode of star formation, namely the single burst galaxies, and those whose star formation has been prolonged, due to either a longer duration or subsequent stellar bursts.

In Figure \ref{fig:hist_alpha_deltat} we show the distributions of the [$\alpha$/Fe] of the two samples. The two distributions are very similar (from a KS-test we get a p-value of 0.89). This suggests that the $\alpha$-enhancement does not have a strong dependence on the SFH, i.e. it does not allow us to distinguish whether the bulk of the stellar population of a galaxy has formed on shorter (< 1 Gyr) or longer (>1 Gyr) time-scales. This is rather unexpected, as the `downsizing' of ETGs predicts higher $\alpha$-enhancement for galaxies that formed faster, as observed in the local Universe (e.g., \citealt{delaRosa+11}). However, the difference in [$\alpha$/Fe] at different SFH may be hidden within our uncertainties. Therefore, we can only conclude that, for massive and quiescent galaxies in LEGA-C, the difference of [$\alpha$/Fe] between galaxies that formed in times shorter and longer than 1 Gyr is less than our typical error, i.e. $\sim0.16$ dex.

Further, we can not evaluate the impact of the SFH on time-scales shorter than the temporal resolution of the MILES models. However, for galaxies with longer star formation, we verified that there is no dependence of [$\alpha$/Fe] on $\Delta\tau'$. This may indicate that their IMF is similar, with mild (if any) correlation with the duration of the star formation. A varying IMF may instead explain the observed variety of [$\alpha$/Fe], but, with these data, we can not disentangle a possible role of the IMF from the uncertainties in our estimates.

We verified that very similar results are obtained when considering different time-scales, like using a different time interval (e.g., $\tau_{90} - \tau_{5}$) or considering the $\tau_{50}$ (or similar proxy for the time at which a galaxy reaches half of its mass), as previous works \citep{delaRosa+11,McDermid+15}.

Finally, as shown in Figure \ref{fig:alpha_mms}, we find no evident correlation between [$\alpha$/Fe] and t$_{\rm form}$ (as defined by equation \eqref{eq:tform}), at least within the uncertainties of our estimates, i.e. there is no significant dependence on the cosmic epoch at which the average stellar mass of galaxies formed.

\section{Probing the evolution of [$\alpha$/Fe] over the last $\sim 6.5$ Gyr}\label{sect:alpha_evol}

In this section, we compare the results on the LEGA-C sample with galaxies at lower redshift. In particular, in section \ref{sect:alpha_local} we compare the [$\alpha$/Fe] values obtained for LEGA-C quiescent galaxies with those of local ETGs from previous studies. Then, for a further comparison, in section \ref{sect:alpha_stack_comp}, we compare the [alpha/Fe] values we derived from stacks of LEGA-C quiescent galaxies with those derived from high-S/N stacks at lower redshifts.

\subsection{Comparing the [$\alpha$/Fe] of LEGA-C quiescients galaxies with local ETGs}\label{sect:alpha_local}

\begin{figure}
\includegraphics[width=\columnwidth]{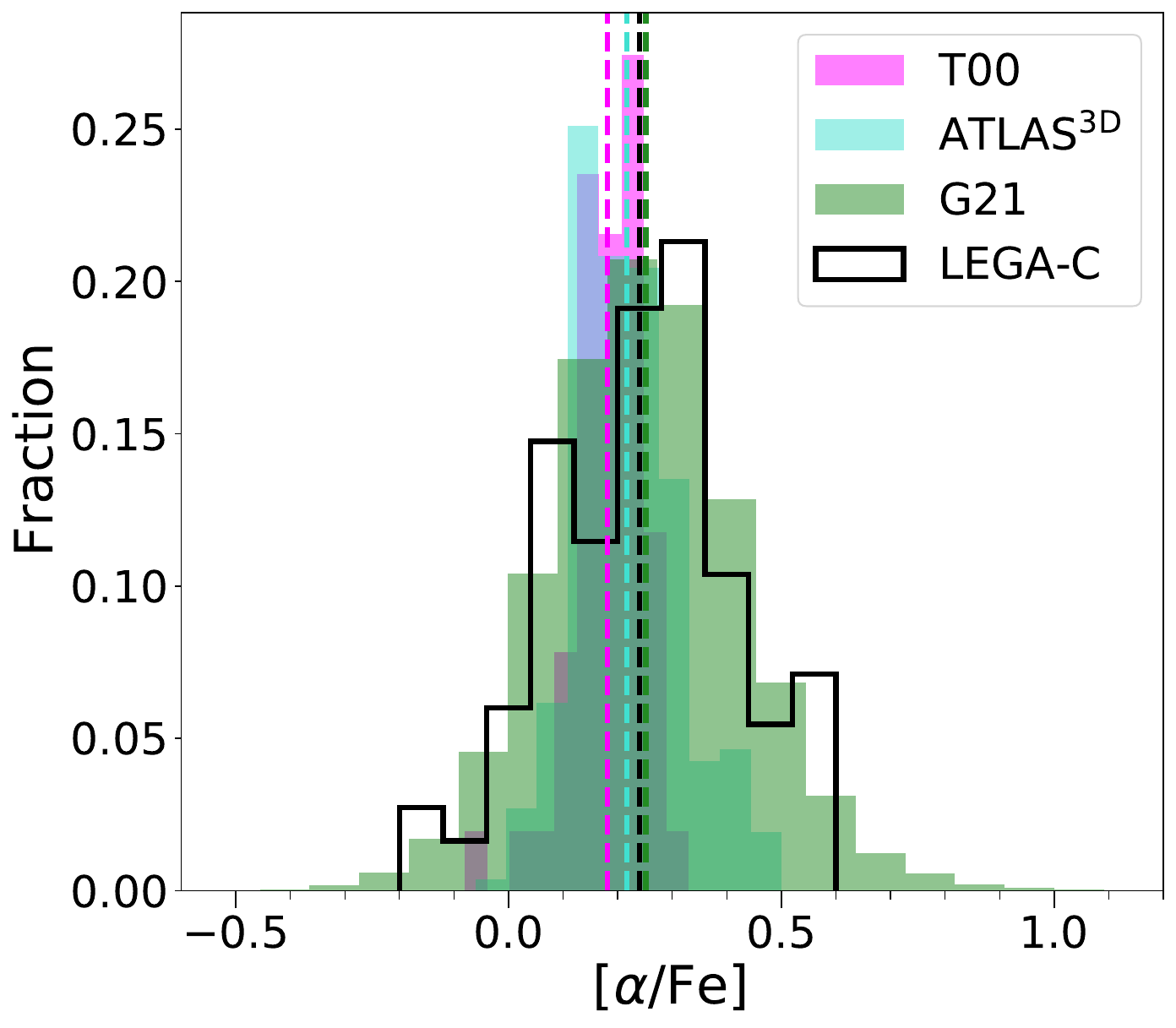}
\caption{Comparison of the distributions of [$\alpha$/Fe] of LEGA-C galaxies (black solid line) with the samples from \protect\cite{Trager+00_1} (T00), \protect\cite{McDermid+15} (ATLAS$^{3\rm D}$) and \protect\cite{Gallazzi+20} (G21). The dashed vertical lines represent the average values of each sample, corresponding to $+0.24\pm0.01$,$+0.19\pm0.01$, $+0.20\pm0.01$, and $+0.25\pm0.00$ dex for LEGA-C, T00, ATLAS$^{3\rm D}$, and G21, respectively.}
\label{fig:alpha_comparison}
\end{figure}

In Figure \ref{fig:alpha_comparison}, we compare the distribution of the [alpha/Fe] values of our LEGA-C sample of quiescent galaxies with the distributions of two samples of local (z<0.02) ETGs, as estimated by \cite{Trager+00_1} (T00) and ATLAS$^{3\rm D}$ \citep{McDermid+15}, and a sample of local quiescent galaxies presented in \cite{Gallazzi+20} (G21), compared to our sample of quiescents from LEGA-C. In particular, differently from this work, quiescent galaxies in G21 have been selected using the specific SFR, and are divided into centrals and satellites (see G21 for details); however, we here ignore this distinction as we verified we get very similar results for the two subsamples. 

In all the three samples from the literature, the [$\alpha$/Fe] values have been estimated by comparing the measured Mg and Fe indices\footnote{More specifically, they all used Mgb to trace the $\alpha$ abundance, while T00 and G21 used Fe5270 and Fe5335, and ATLAS$^{3\rm D}$ used Fe5015 and Fe5270 to trace the iron abundance.} with those predicted by SSP models. However, data, models, and methods to estimate the [$\alpha$/Fe] values are independent. 

The distributions of T00 and ATLAS$^{3\rm D}$ are similar, being gaussianly distributed at [$\alpha$/Fe] = $+0.19\pm0.01$ dex and $+0.20\pm0.01$ dex with standard deviations of 0.07 and 0.10, respectively. The sample of G21, instead, is gaussianly distributed at a higher average [$\alpha$/Fe] $= 0.25\pm0.00$ dex, and with a larger standard deviation (0.18), comparable with the one of the LEGA-C sample (0.17). The large errors on the [$\alpha$/Fe] of LEGA-C galaxies do not allow us to establish whether the scatter may be the result of cosmic evolution or, rather, of the uncertainty in the measurements themselves. However, the agreement with the distribution of G21 sample of passive galaxies suggests that the evolution did not play a significant role. Overall, we can safely say that the average values of [$\alpha$/Fe] of all the three samples of local ETGs or quiescent galaxies are close to the [$\alpha$/Fe]$= +0.24$ dex estimated for LEGA-C galaxies.

We thus conclude that the average [$\alpha$/Fe] has not changed from $z=0.75$ to $z=0$.

\begin{table*}
	\centering
	\caption{Table of properties of stacked galaxies.}
	\label{tab:mgfe}
	\begin{center}
	
	\begin{tabular}{ccccccc} 
		\hline
		Name & N$_{\rm spec}$ & <log$_{10}$M$_*$> & <$\sigma_*$> & Fe4383 & Mgb & [$\alpha$/Fe]\\
		& & (M$_\odot$) & (km s$^{-1}$) & & & (dex)\\
		(1) & (2) & (3) & (4) & (5) & (6) & (7)\\
		\hline
		SR+20 & 13 & $11.3$ & 250 & $3.94\pm0.16$ & $3.41\pm0.17$ & $+0.23\pm0.00$\\
		SPIDER & 8 & 10.7 & 150 & $4.97\pm0.06$ & $3.92\pm0.12$ & $+0.14\pm0.00$\\
		SPIDER & 8 & 11.1 & 250 & $4.98\pm0.04$ & $4.35\pm0.05$ &  $+0.23\pm0.00$\\
		LEGA-C & 21 & $10.5$ & 165 & $3.68\pm0.83$ & $3.40\pm0.33$ & $+0.28\pm0.02$\\
		LEGA-C & 42 & $10.7$ & 168 & $3.79\pm0.69$ & $3.73\pm0.47$ & $+0.34\pm0.02$\\
		LEGA-C & 37 & $10.9$	& 204 & $4.13\pm0.28$ & $3.62\pm0.32$ & $+0.24\pm0.02$\\
		LEGA-C & 41 & $11.1$ & 220 & $4.39\pm0.42$ & $3.73\pm0.38$ & $+0.22\pm0.02$\\
		LEGA-C & 21 & $11.3$ & 242 & $4.55\pm0.36$ & $3.86\pm0.17$ & $+0.22\pm0.03$\\
		\hline
	\end{tabular}
		\end{center}

	\begin{minipage}{\linewidth}
	Columns: (1) Name corresponding to the values plotted in Figure \ref{fig:mgfe}. (2) Number of spectra over which indices are estimated. (3) Average stellar mass of the N$_{\rm spec}$ spectra. (4) Average stellar velocity dispersion of the N$_{\rm spec}$ spectra (5) Estimated Fe4383 and associated error (6) Estimated Mgb and associated error. (7) Estimated [$\alpha$/Fe] and associated error from the Mgb and Fe4383 indices (section \ref{sect:alpha_stack}), calculated from values provided in column (5) and (6); the uncertainties are the gaussian errors on the mean. The median redshifts of LEGA-C, SR+20, and SPIDER are $z = 0.68, 0.38$, and 0.07, respectively. For SR+20 and SPIDER galaxies, the values of indices reported in this table are the average values of the single measurements of the N$_{\rm spec}$ stacked spectra, and the errors are the corresponding median absolute deviations. Instead, the reported values of LEGA-C stacks are the values we measure with LECTOR on the stacked spectra, while the errors are the median absolute deviations of the single galaxies' values constituting the stacked spectrum. 
	\end{minipage}

\end{table*}

\subsection{Comparing stack of LEGA-C quiescents with stack of quiescients at lower redshifts}\label{sect:alpha_stack_comp}

To consolidate the results obtained in the previous section, we further compare LEGA-C quiescent galaxies with high S/N spectra of \cite{SR+20} (SR+20) (13 stacked spectra) and \cite{spider} (SPIDER) (16 stacked spectra). The spectra of both studies are high S/N ($\sim 100$) stacked spectra of ETGs observed at median redshifts $z = 0.38$ and $z = 0.07$, respectively. The range in velocity dispersion covered by SPIDER galaxies is similar to that of LEGA-C galaxies, namely $\sigma_* = 100 - 320$ km s$^{-1}$, and the two samples have a comparable mass range $\log_{10}$(M$_*$/M$_\odot) = 10.6 - 11.2$. On the other hand, galaxies from SR+20 have slightly different but comparable velocity dispersions, $\sigma_* = 160 - 340$ km s$^{-1}$, and larger masses, $\log_{10}$(M$_*$/M$_\odot) = 11.2 - 11.5$. 

\subsubsection{Mgb vs Fe4383 diagram}\label{sect:mgfe_evol}

We measure the Fe4383 and Mgb indices of each stacked spectrum of SR+20 and SPIDER using LECTOR. For SR+20 we take the average measurements of the 13 stacks, corresponding to an average mass $\log_{10}$(M$_*$/M$_\odot) \approx 11.3$, and an average velocity dispersion $\sigma_* = 250$ km s$^{-1}$. For SPIDER galaxies, we split the sample into two, and average the spectral indices of the stacked spectra with masses $10.6 \lesssim \log_{10}$(M$_*$/M$_\odot ) \lesssim 10.9$ and $10.9 \lesssim \log_{10}$(M$_*$/M$_\odot ) < 11.2$, whose average velocity dispersions are 150 and 250 km s$^{-1}$, respectively. In Table \ref{tab:mgfe} we summarize the relevant properties of the SR+20, SPIDER, and LEGA-C stacked spectra.

For a proper comparison, we stack spectra of LEGA-C galaxies\footnote{To perform the stacking, the spectra have been first shifted to the rest-frame, then normalized to the mean flux measured in the rest-frame wavelength $4000 - 4100$ \AA , and re-sampled to a common dispersion of 1 \AA \, pixel$^{-1}$. Finally, we considered the median value of the fluxes of the stacked galaxies as the flux of the stacked spectrum and the median absolute deviation as the associated error.} at five different mass bins, ranging from $\log_{10}$(M$_*$/M$_\odot) = 10.4$ to $11.4$, at a step of 0.2 dex, and measure the indices with LECTOR. As the errors, we consider the median absolute deviations of the measurements of single galaxies' constituting the stacked spectrum. We summarize the relevant properties of LEGA-C stacks in Table \ref{tab:mgfe}.

The indices derived for the stacked spectra of LEGA-C galaxies, as well as SPIDER and SR+20, are plotted in Figure \ref{fig:mgfe}. SPIDER galaxies are pretty much aligned (i.e. they lie close to the same line of the grid) to the LEGA-C stacks. The fact that SPIDER galaxies, observed at $z\approx 0.07$, have higher indices than LEGA-C stacked galaxies, observed at $z\approx 0.7$, is likely due to the older ages of the former. Indeed, as pointed out above, when comparing the grids at 3.5 and 6.5 Gyr, both indices increase parallel to metallicity at older ages. 

To quantify the effect of a passive evolution on the indices from $z\sim0.7$ to $z\sim0$, we estimate the variation of Mgb and Fe4383 purely due to aging, similarly to section \ref{sect:agemet}. More specifically, we measure the indices of the E-MILES model with age 3.5 Gyr and metallicity [M/H] $= + 0.15$ (i.e., similar to the median values estimated for the LEGA-C galaxies, see section \ref{sect:agemet}). Also, we measure the indices of the model with age 9.5 Gyr and the same metallicity; namely, we are considering the change in the indices purely due to aging of 6 Gyr (i.e., from $z\sim0.7$ to $z\sim0$). For Mgb and Fe4383 we estimate a percentage variation of $16\%$, and $19\%$, respectively. This is comparable with the relative difference in the measured indices between the SPIDERs and LEGA-C stacks, at similar masses ($\sim 11 \%$ and $16 \%$, compare with table \ref{tab:mgfe}). We thus conclude that the age difference between the LEGA-C and SPIDER galaxies can account for the difference in the measured indices.

Compared to SPIDER and LEGA-C galaxies, the behavior of SR+20, in Figure \ref{fig:mgfe}, is more difficult to explain. Indeed, although the point is aligned to both LEGA-C and SPIDER galaxies, the index line strengths are lower compared to those for the LEGA-C or SPIDER with similar stellar masses (i.e. comparing those with average log$_{10}$(M$_*$/M$_\odot$) = 11.1 - 11.3). Although we can not explain this behavior, which might reflect the different selection of the SR+20 with respect to the SPIDER and LEGA-C galaxies (see \citealt{SR+20} for details), SR+20 galaxies still have supersolar [$\alpha$/Fe], with values comparable with LEGA-C galaxies. We investigate these results, more quantitatively, in the following section.

\subsubsection{[$\alpha$/Fe] estimates}\label{sect:alpha_stack}

Using the same method of section \ref{sect:alpha_estimate}, we estimate the $\alpha$-enhancement of the LEGA-C, SR+20, and SPIDER stacked spectra. In particular, for SR+20, and SPIDER, we estimate [$\alpha$/Fe] from the same average values of Mgb and Fe4383, and corresponding errors, tabulated in Table \ref{tab:mgfe}. Note that we should in principle use grids of older ages for SR+20 and SPIDER; however, we have already highlighted (section \ref{sect:mgfe} and \ref{sect:mgfe_evol}) that grids of older ages would provide virtually the same [$\alpha$/Fe] estimates.

The values of [$\alpha$/Fe] estimated for the stacked spectra are given in Table \ref{tab:mgfe}. As already pointed out, the spectral indices of stacked galaxies from LEGA-C, SPIDER, and SR+20 lie close to the same line of the grid, and indeed their values of [$\alpha$/Fe] are very similar, and most cases are close to [$\alpha$/Fe]$\sim 0.2$. In particular, all LEGA-C stacks, SPIDER and SR+20 with masses $\geq10^{11}$ M$_\odot$ have a remarkably similar $\alpha$-enhancement, $\approx +0.23$ dex (which is also close to the average value estimated for the whole LEGA-C sample), notwithstanding the rather large redshift range covered by these data, corresponding to almost 6.5 Gyr of cosmic time.

This confirms the result that, \textit{the [$\alpha$/Fe] of passive galaxies has not changed systematically and significantly over the last $\sim 6.5$Gyr, from $z=0.75$ to $z=0$.}

\section{Summary and Conclusions}\label{sect:summary}

In this work, we have estimated and studied the [$\alpha$/Fe] of a sample of 183 quiescent galaxies, selected from the LEGA-C survey, at redshift $0.60 \leq z \leq 0.75$. In particular, we used Mgb as a proxy for the $\alpha$ elements, and Fe4383 for the iron abundance. We have derived the [$\alpha$/Fe] by comparing the observed indices with those predicted by MILES models.

The summary of our results is the following:

\begin{enumerate}[(i)]

\item The distribution of [$\alpha$/Fe] of LEGA-C quiescent galaxies (Figure \ref{fig:hist_alpha}) is approximately gaussian, with an average value of [$\alpha$/Fe] $= 0.24 \pm 0.01$ dex. In particular, $91\%$ of galaxies in our sample have super-solar $\alpha$ abundance, while the remaining $9\%$ is alpha-depleted or solar-scaled.\\

\item The spectral indices show a slight increase with the global metallicity, the stellar mass, and the stellar velocity dispersion (Figure \ref{fig:mgfe}). However, no trend is spotted with the [$\alpha$/Fe] estimates, even when considering galaxies with higher S/N, but it may be hidden within the large uncertainties (Figure \ref{fig:alpha_mms}).\\

\item The distributions of [$\alpha$/Fe] values are similar, within the typical uncertainty of 0.16 dex, for galaxies that formed the bulk of their stellar mass in a time shorter and longer than 1 Gyr (Figure \ref{fig:hist_alpha_deltat}). This suggests that, outside the temporal resolution of the models adopted (0.5 Gyr), the impact of SFH on the [$\alpha$/Fe] of galaxies should be milder than our estimated uncertainty (0.16 dex). Thus, it can not account for the whole distribution of [$\alpha$/Fe] values. Additionally, we find no correlation of [$\alpha$/Fe] with the galaxies' formation time.\\

\item We compare the distributions of [$\alpha$/Fe] of LEGA-C galaxies with those of local ETGs. We find no significant difference in the average [$\alpha$/Fe]$\approx +0.2$ dex (Figure \ref{fig:alpha_comparison}). This result is confirmed by comparing the stacked spectra of LEGA-C galaxies with high-S/N stacked spectra from SR+20 and SPIDER galaxies, at $z=0.38$ and $z=0.07$. Indeed, we find very similar values of [$\alpha$/Fe], especially for masses $\geq 10^{11}$M$_\odot$, for which all have [$\alpha$/Fe] $= 0.22 - 0.24$ dex (section \ref{sect:alpha_stack}). This suggests a lack of evolution in the average [$\alpha$/Fe]($\approx +0.2$ dex) over the last $\sim 6.5$ Gyr of the Universe.

\end{enumerate}

The great majority of quiescent galaxies at intermediate redshift, $0.60 \leq z \leq 0.75$, are $\alpha$-enhanced, like those at $z\approx0$. The correlations of [$\alpha$/Fe] with mass and $\sigma_*$ observed in the local Universe may persist, but we could not recover them, due to large uncertainties. 

Within the errors, we do not see significant differences when comparing galaxies that formed most of their mass before and after 1 Gyr. Hence, the SFH should behave in such a way that, outside the temporal resolution of the models, it does not affect the distribution of [$\alpha$/Fe] more significantly than the typical uncertainty, 0.16 dex. Additionally, the time at which galaxies form does not seem to play an important role. This implies that if the IMF plays a role in determining the [$\alpha$/Fe] of a galaxy, this is within the earliest times of its formation ($\Delta\tau' < 1$ Gyr), and independently of the redshift at which it formed. However, varying IMF, as well as varying time scales of the star formation shorter than 1 Gyr, may play a role in determining the overall distribution of [$\alpha$/Fe] values.

The fact that the mean values estimated for the SR+20 and SPIDER galaxies, as well as the distributions of [$\alpha$/Fe] values of local galaxies, are remarkably consistent with those of LEGA-C galaxies suggests that the cosmic evolution has not altered the average [$\alpha$/Fe], from $z=0.75$ to $z=0$.

All these results indicate that time, intended as formation epoch, and star formation time scales, plays a marginal role in determining the overall distribution of [$\alpha$/Fe] values of quiescent galaxies, and that cosmic evolution did not affect \textit{significantly and systematically} these values, at least in the last $\sim 6.5$ Gyr of the Universe. If it did, the effects are buried within the observational errors.

\section*{Acknowledgements}

D.B., P.S., F.L.B., R.D.P., A.G., A.P., and C.S. acknowledge support by the grant PRIN-INAF-2019 1.05.01.85. A.G. acknowledges support from INAF-Minigrant-2022 "LEGA-C" 1.05.12.04.01. C.S. is supported by a `Hintze Fellowship' at the Oxford Centre for Astrophysical Surveys, which is funded through generous support from the Hintze Family Charitable  Foundation. 

\section*{Data Availability}

The spectra of LEGA-C galaxies used in this analysis, as well as the estimates of the spectral indices and relative errors, are taken from the DR3 of the LEGA-C survey, and are publicly available. The values tabulated in Table \ref{tab:alpha} are provided in electronic format as supplementary material to this paper.



\bibliographystyle{mnras}
\bibliography{biblio}



\appendix

\section{Estimates of [$\alpha$/Fe] using additional iron lines}\label{app:fe}

\begin{figure}
\includegraphics[width=\columnwidth]{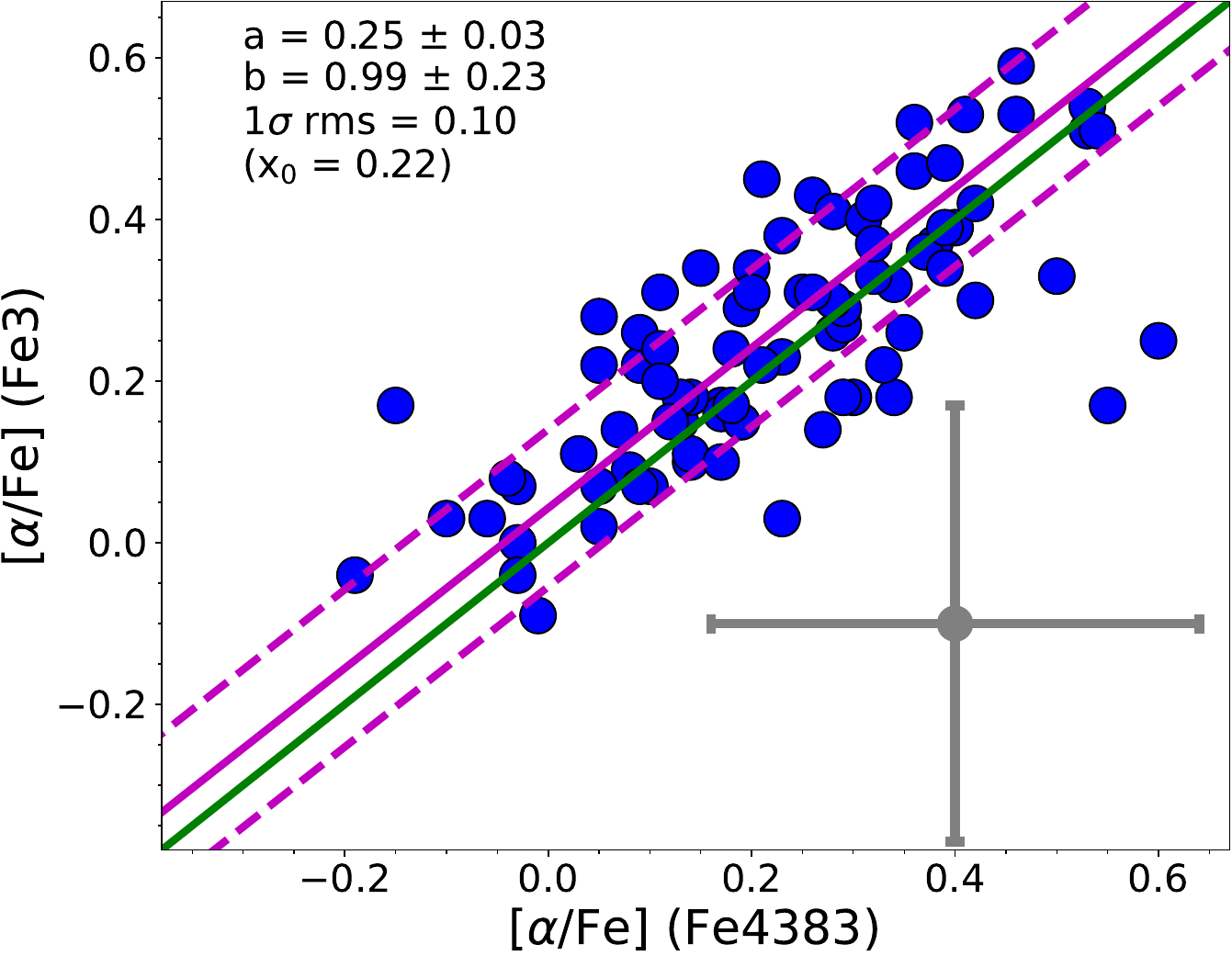}
\caption{Comparison between the [$\alpha$/Fe] estimates using Fe4383 and Fe3. The solid magenta line is the fit of the linear relation $y = a + b (x - x_0)$, with $x_0$ being the median [$\alpha$/Fe] estimated from Fe4383. The dashed magenta lines are the $\pm 1\sigma$ scatter of the linear fit. The solid green line is the one-to-one relation. The grey point indicates the average uncertainty of the [$\alpha$/Fe] estimates.}
\label{fig:fe4383_fe3_direct}
\end{figure}

As discussed in section \ref{sect:data}, we do not use the Fe4531 index, as it is noisier than Fe4383 (the relative errors are typically $\sim 40\%$ larger) and would lead to an increase of the scatter on [$\alpha$/Fe] abundance estimates. Similarly, we do not use Fe5406, given that the relative errors on this index are typically $\sim 70\%$ larger than Fe4383; moreover, Fe5406 is measured for less than a third of the galaxies in our sample. We do not use the Fe5015 line since it may be contaminated by the [OIII] emission line at 5007 \AA . Finally, even though the Fe5270 and Fe5335 lines have comparable uncertainties to Fe4383 and would improve the accuracy of our estimates, we do not use them since they are measured for less than half of the galaxies in our sample (83/183). However, we can use these indices to asses how the estimates of [$\alpha$/Fe] would change using a different proxy for iron abundance.

To this aim, we use a combination of Fe4384, Fe5270, and Fe5335 as a proxy for iron abundance and repeat the analysis to estimate [$\alpha$/Fe] (section \ref{sect:alpha/fe}). Specifically, we define Fe3 = (Fe4384 + Fe5270 + Fe5335)/3, construct a grid of Mgb and Fe3 from MILES models convolved by gaussians with width $\sigma_* = 250$ km s$^{-1}$ at different [$\alpha$/Fe] and [M/H] values, and compare it with the measured indices.

In Figure \ref{fig:fe4383_fe3_direct} we compare the [$\alpha$/Fe] values obtained from Fe3 and Fe4383 for the 83 galaxies with both estimates available. We fit the linear relation $y = a + b (x - x_0)$, with $x_0$ being the median [$\alpha$/Fe] estimated from Fe4383, and compare it with the one-to-one relation (green solid line). The comparison indicates an average offset of 0.03 dex towards higher [$\alpha$/Fe] values when using Fe3. However, the observed scatter is significantly larger ($\sim 0.1$ dex), due to the large uncertainties in index measurements, and the [$\alpha$/Fe] estimates are consistent for most galaxies. With only two exceptions, all estimates are consistent within the errors.

\begin{figure}
\center
\includegraphics[width=.7\columnwidth]{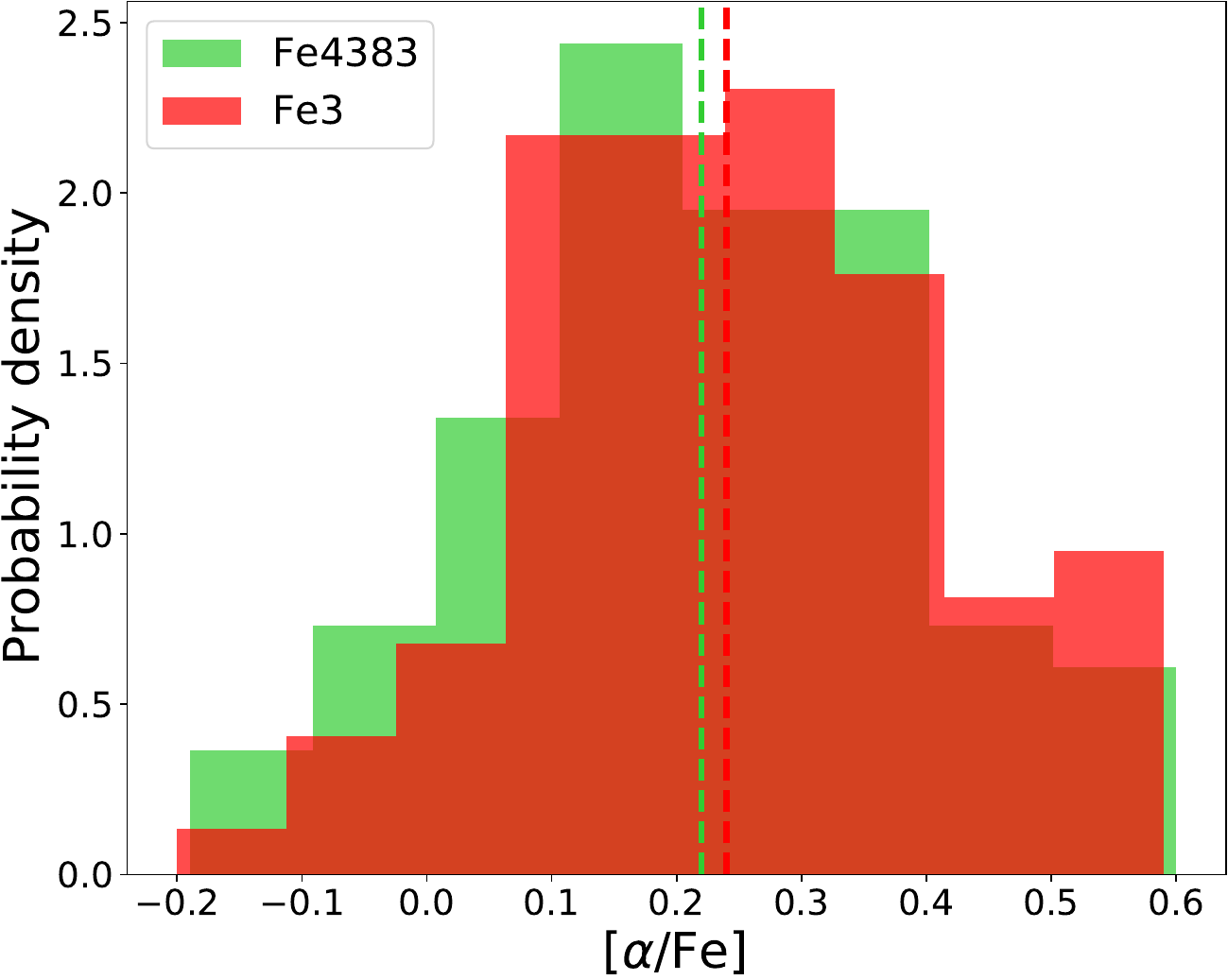}
\caption{Distributions of [$\alpha$/Fe] estimated using Fe4383 (green) and Fe3 (red). The corresponding average values, plotted with dashed lines, are $0.22\pm0.02$ and $0.24\pm0.02$, respectively.}
\label{fig:fe3_fe4383}
\end{figure}

In Figure \ref{fig:fe3_fe4383} we further compare the histograms of [$\alpha$/Fe] obtained from Fe3 and Fe4383. We find an average [$\alpha$/Fe] = $0.24\pm0.02$ when using the Fe3. Note that, for this subsample, the average [$\alpha$/Fe] = $0.22\pm0.02$ when using only Fe4383. The average values of the two samples are consistent within the errors at 1$\sigma$, and the KS-test indicates that the two distributions are consistent, with a p-value of 0.72.

We conclude that using a different proxy for iron abundance does not change our results.

\bsp	
\label{lastpage}
\end{document}